\documentclass[12pt]{article}

\usepackage{natbib,graphicx}

\begin{document}

\title{Lyapunov and diffusion timescales in the solar neighborhood}

\author{Ivan I. Shevchenko\thanks{E-mail:~iis@gao.spb.ru} \\
Pulkovo Observatory of the Russian Academy of Sciences, \\
Pulkovskoje ave. 65, St.Petersburg 196140, Russia}

\maketitle

\begin{abstract}
We estimate the Lyapunov times (characteristic times of
predictability of motion) in Quillen's models for the dynamics in
the solar neighborhood. These models take into account
perturbations due to the Galactic bar and spiral arms. For
estimating the Lyapunov times, an approach based on the separatrix
map theory is used. The Lyapunov times turn out to be typically of
the order of 10 Galactic years. We show that only in a narrow
range of possible values of the problem parameters the Galactic
chaos is adiabatic; usually it is not slow. We also estimate the
characteristic diffusion times in the chaotic domain. In a number
of models, the diffusion times turn out to be small enough to
permit migration of the Sun from the inner regions of the Milky
Way to its current location. Moreover, due to the possibility of
ballistic flights inside the chaotic layer, the chaotic mixing
might be even far more effective and quicker than in the case of
normal diffusion. This confirms the dynamical possibility of
Minchev and Famaey's migration concept.

\end{abstract}

\section{Introduction}

Chaotic dynamics due to interaction of nonlinear resonances in
Hamiltonian systems is studied in a broad range of application
areas, from plasma physics to celestial mechanics (e.g.,
\citealt{C79,LL92,A06}). The characteristic time of predictability
of any motion is nothing but the Lyapunov time (the inverse of the
maximum Lyapunov exponent) of the motion. Generally, the
estimation of the Lyapunov exponents is one of the most important
tools in the study of chaotic motion \citep{LL92}, in particular
in celestial mechanics. The Lyapunov exponents characterize the
mean rate of exponential divergence of trajectories close to each
other in phase space. Non-zero Lyapunov exponents indicate chaotic
character of motion, while the maximum Lyapunov exponent equal to
zero signifies regular (periodic or quasiperiodic) motion.

The development of methods of numerical computation of the
Lyapunov exponents has more than a 30 year history (e.g.,
\citealt{F84,LL92}). In contrast, methods of analytical estimation
of the Lyapunov exponents started to be developed only recently
\citep{HM96,MH97,S00a,S02,S07,S08b}.

In studies of the dynamics of the Milky Way, analysis of the
Lyapunov exponents has not yet been widely used, but nevertheless
there are important achievements. \cite{F01} used Lyapunov
exponents as a tool to find bar-induced manifestations of chaos in
the local disk stellar kinematics. Taking into account the
perturbations from the bar solely (for some particular bar
strengths), he constructed ``Lyapunov diagrams'', presenting the
Lyapunov timescales of the orbits in the $u$~--~$v$ velocity plane
at fixed space positions, and identified regular and chaotic
domains in velocity space as a function of space position with
respect to the bar. The Lyapunov exponents were calculated on a
Cartesian grid of planar velocities. The fraction of chaotic
orbits was demonstrated to obviously increase with the bar
strength. However, the diagrams in \cite{F01} can hardly be used
to estimate real values of Lyapunov times, because the saturation
of the computed values of the Lyapunov exponents was not
controlled, while the adopted computation time, corresponding to
three Hubble times (equivalently, $\sim 100$ Galactic years),
might not be enough for the saturation. In other words, the
obtained values characterize not the whole chaotic regions of the
phase space, but only rather small vicinities of the initial data.
Therefore, the computed values are the ``local'' values of the
Lyapunov exponents. This is testified by the fact that the
variation of the Lyapunov exponents in the diagrams of \cite{F01}
is smooth, while there must be a sharp distinction between the
chaotic regions (with non-zero Lyapunov exponents) and regular
regions (with zero Lyapunov exponents) in the divided phase space.

In connection with the problem of estimation of Lyapunov
timescales in the solar neighborhood, \cite{Q03} noted that,
according to \cite{HM96}, for a fully overlapped system, the
chaotic zone should have a Lyapunov time $\sim 2 \pi / \nu$ (where
$\nu$ is the frequency of perturbation), corresponding to the
separatrix pulsation period. In what follows we shall consider the
model set of \cite{Q03} for the stellar dynamics in the solar
neighborhood and show that the heuristic estimate $\sim 2 \pi /
\nu$ severely underestimates the real Lyapunov time. Besides, we
shall see that it is rather seldom that the considered dynamical
systems, modeling the dynamics in the solar neighborhood, can be
called ``fully overlapped''.

Besides obtaining the Lyapunov times, we shall estimate the
diffusion times in the chaotic domain of the phase space, in the
same model set. This will allow one to judge on the possibility
for migration of the Sun from the inner regions of the Milky Way
to its current location. Such a possibility, arising due to the
overlapping of resonances in the phase space, was advocated and
studied in detail by \cite{MF10} and \cite{MFC11}.

\section{The model of interaction of nonlinear resonances}
\label{minr}

Many problems on nonlinear resonances in mechanics and physics are
described by the perturbed pendulum Hamiltonian

\vspace{-3mm}

\begin{equation}
H = {{{\cal G} p^2} \over 2} - {\cal F} \cos \varphi +
    a \cos(\varphi - \tau) + b \cos(\varphi + \tau)
\label{h}
\end{equation}

\noindent (e.g., \citealt{S00b}). The first two terms in
equation~(\ref{h}) represent the Hamiltonian $H_0$ of the
unperturbed pendulum; $\varphi$ is the pendulum angle (the
resonance phase angle) and $p$ is the momentum. The periodic
perturbations are given by the last two terms; $\tau$ is the phase
angle of perturbation: $\tau = \Omega t + \tau_0$, where $\Omega$
is the perturbation frequency and $\tau_0$ is the initial phase of
the perturbation. The quantities ${\cal F}$, ${\cal G}$, $a$, $b$
are parameters.

Generally, equation~(\ref{h}) describes a triplet (triad) of
resonances: there are three trigonometric terms, each
corresponding to a particular resonance. In the following
Sections, the case of a resonance duad is considered (i.e., $a$ or
$b$ equals zero).

It is convenient to describe the motion in the vicinity of the
separatrices of the Hamiltonian~(\ref{h}) by means of the
so-called separatrix map \citep{C79}. It is a two-dimensional,
area-preserving map given by

\vspace{-3mm}

\begin{eqnarray}
& & w_{i+1} = w_i - W \sin \tau_i,  \nonumber \\
& & \tau_{i+1} = \tau_i +
                 \lambda \ln {32 \over \vert w_{i+1} \vert}
                 \ \ \ (\mbox{mod } 2 \pi) .
\label{sm}
\end{eqnarray}

\noindent These equations give the classical separatrix map
\citep{C79}, valid in the symmetric ($a = b$) case of
perturbation. The variable $w$ of the map denotes the relative
(with respect to the unperturbed separatrix value) pendulum energy
$w \equiv {H_0 \over {\cal F}} - 1$, and $\tau$ retains its
meaning of the phase angle of perturbation. The parameter
$\lambda$ is the ratio of $| \Omega |$, the absolute value of the
perturbation frequency, to $\omega_0 = | {\cal F G} |^{1/2}$, the
frequency of the small-amplitude pendulum oscillations. The
parameter $W$ in the case of $a = b$ has the form~\citep{S98}

\vspace{-3mm}

\begin{equation}
W = \varepsilon \lambda \left( A_2(\lambda) + A_2(-\lambda)
\right) = {4 \pi \varepsilon \lambda^2 \over \sinh {\pi \lambda
\over 2}} ,
\label{W1}
\end{equation}

\noindent where $\varepsilon = a / {\cal F}$, and $A_2(\lambda)$
is the value of the Melnikov--Arnold integral as defined in
\cite{C79}:

\begin{equation}
A_2(\lambda) = 4 \pi \lambda {\exp({{\pi \lambda} / 2}) \over
\sinh (\pi \lambda)}.
\label{A2}
\end{equation}

\noindent Formula~(\ref{W1}) differs from that given in \cite{C79}
and \cite{LL92} by the term $A_2(-\lambda)$, which is small for
$\lambda \gg 1$. However, its contribution is significant for
small values of $\lambda$~\citep{S98}, i.e., in the case of
adiabatic chaos. The kind of chaos (adiabatic or non-adiabatic) in
model~(\ref{h}) is identified by the value of $\lambda$: if
$\lambda < 1/2$, it is slow (adiabatic), otherwise it is ``fast''
(non-adiabatic; \citealt{S08a}).

One iteration of the separatrix map corresponds to one period of
the pendulum rotation or a half-period of its libration. The
applicability of the theory of separatrix maps for description of
the motion near the separatrices of the perturbed nonlinear
resonance in the full range of the relative frequency of
perturbation, including its low values, was considered and shown
to be legitimate in \cite{S00b}.

The separatrix map in the case of asymmetric ($a \neq b$)
perturbation is different from that in the symmetric ($a = b$)
case, because the energy increments are different for the prograde
and the retrograde motions of the model pendulum \citep{S99}. (The
motion is called prograde or retrograde if the variation of
$\varphi$ with time is, respectively, positive or negative.) The
algorithm, taking this difference into account, constitutes the
separatrix algorithmic map \citep{S99}:

\vspace{-3mm}

\begin{eqnarray}
& & \mbox{if } w_i < 0 \mbox{ and } W = W^-
              \mbox{ then } W := W^+, \nonumber \\
& & \mbox{if } w_i < 0 \mbox{ and } W = W^+
              \mbox{ then } W := W^-, \nonumber \\
& & w_{i+1} = w_i - W \sin \tau_i,   \nonumber \\
& & \tau_{i+1} = \tau_i + \lambda \ln {32 \over \vert w_{i+1} \vert}
                   \ \ \ (\mbox{mod } 2 \pi),
\label{sam}
\end{eqnarray}

\noindent where $W^+$ and $W^-$ denote the values of the $W$
parameter for the prograde and retrograde motions respectively. In
the case of asymmetric perturbation these values are different.

Equations~(\ref{sam}) as well can be written in a shorter way
\citep{S00b}:

\vspace{-3mm}

\begin{eqnarray}
& & \mbox{if } w_i < 0 \mbox{ and } W = W^\pm
              \mbox{ then } W := W^\mp, \nonumber \\
& & w_{i+1} = w_i - W \sin \tau_i,   \nonumber \\
& & \tau_{i+1} = \tau_i + \lambda \ln {32 \over \vert w_{i+1} \vert}
                   \ \ \ (\mbox{mod } 2 \pi).
\label{sam1}
\end{eqnarray}

\noindent The sign in the upper index of $W$ alternates at each
iteration if $w_i < 0$ (i.e., at librations); the quantity $W^\pm$
denotes $W^+$ or $W^-$, while $W^\mp$ denotes a corresponding
value of $W^-$ or $W^+$.

The essence of the separatrix algorithmic map is in taking into
account the alternations of the $W$ parameter. It alternates when
the direction of motion alternates. This takes place either when
rotation changes to libration, or when the motion is librational.
Algorithms~(\ref{sam}) and (\ref{sam1}) do not contain the
condition $w_i > 0$, because the direction of motion does not
change when it holds.

In order to find expressions for $W^+$ and $W^-$, one should
integrate the increment of energy per one iteration of the map,
following the usual procedure \citep{C79}, but making it
separately for prograde and retrograde directions of motion. This
gives \citep{S99}

\vspace{-3mm}

\begin{equation}
 W^+ (\lambda, \eta) = \varepsilon
 \lambda \left( A_{2}(\lambda) + \eta A_{2}(-\lambda) \right),
\label{Wplus}
\end{equation}

\vspace{-3mm}

\begin{equation}
 W^- (\lambda, \eta) = \varepsilon
 \lambda \left( \eta A_{2}(\lambda) + A_{2}(-\lambda) \right),
\label{Wmin}
\end{equation}

\noindent where $\varepsilon = a / {\cal F}$ and $\eta = b / a$.
The Melnikov--Arnold integral $A_2(\lambda)$ is given by
equation~(\ref{A2}).

The separatrix map theory can be used for analytical estimation of
the maximum Lyapunov exponents \citep{S00a,S02,S07,S08b}.
Comparisons of the predictions of this theory  with
numerical-experimental results can be found in \cite{SK02} and
\cite{S00a,S02,S07,S08b,S09}, where it was applied to various
problems of celestial mechanics: rotational dynamics of planetary
satellites, orbital dynamics of satellite systems, and orbital
dynamics of asteroids.

The maximum Lyapunov exponent is defined by the limit

\begin {equation}
L = \limsup _ {{t \to \infty} \atop {d (t_0) \to 0}} {1 \over
{t-t_0}} \ln {d (t) \over d (t_0)} ,
\label {def_lce}
\end {equation}

\noindent where $d(t_0)$ is the distance (in phase space) between
two nearby initial conditions for two trajectories at the initial
moment of time $t_0$, $d(t)$ is the distance between the evolved
initial conditions at time $t$ (e.g., \citealt{LL92}).

According to the general approach used by \cite{S00a,S02,S07}, the
maximum Lyapunov exponent, $L$, of the motion in the main chaotic
layer of system given by Eq.~(\ref{h}) is represented as the ratio
of the maximum Lyapunov exponent, $L_{\mathrm{sx}}$, of its
separatrix map and the average period $T$ of rotation (or,
equivalently, the average half-period of libration) of the
resonance phase angle $\varphi$ inside the layer: $L =
L_{\mathrm{sx}}/T$. In this way, formulas for the Lyapunov time
were derived in \cite{S07} for four generic cases of interacting
resonances: the fastly chaotic resonance triplet, fastly chaotic
resonance doublet, slowly chaotic resonance triplet, and slowly
chaotic resonance doublet (called, respectively, ``ft,'' ``fd,''
``st,'' and ``sd'' resonance multiplet types).

In the following Sections, we shall need formulas for the Lyapunov
times for the ``fd'' and ``sd'' resonance types only. The formula
for the Lyapunov time for the ``fd'' resonance type is

\begin{equation}
T_{\mathrm L} = \frac{T_{\mathrm{pert}}}{2 \pi} \cdot
\frac{\mu_{\mathrm{libr}} + 1} { \mu_{\mathrm{libr}}
\frac{L_{\mathrm{sx}}(2 \lambda)}{T_{\mathrm{sx}}(2 \lambda, W)} +
\frac{L_{\mathrm{sx}}(\lambda)}{T_{\mathrm{sx}}(\lambda, W)}},
\label{TLfd}
\end{equation}

\noindent where $\mu_{\mathrm{libr}} \approx 4$ \citep{S07}. The
quantity $T_{\mathrm{pert}} = 2 \pi / | \Omega |$ is the period of
perturbation. The quantities $W$, $L_{\mathrm{sx}}$, and
$T_{\mathrm{sx}}$ are given by the formulas

\begin{equation}
W(\varepsilon, \lambda) =
\varepsilon \lambda \left( A_2(\lambda) + A_2(-\lambda) \right) =
{4 \pi \varepsilon \lambda^2 \over \sinh {\pi \lambda \over 2}},
\label{W}
\end{equation}

\begin{equation}
L_{\mathrm{sx}}(\lambda) \approx C_h {2 \lambda \over 1 + 2
\lambda},
\label{Lsx}
\end{equation}

\begin{equation}
T_{\mathrm{sx}}(\lambda, W) \approx \lambda \ln {32 e \over
\lambda | W |},
\label{Tsx}
\end{equation}

\noindent where $C_h \approx 0.80$ is Chirikov's constant
\citep{S04} and $e$ is the base of natural logarithms.

The formula for the Lyapunov time for the ``sd'' resonance type is

\begin{equation}
T_{\mathrm L} \approx  {T_{\mathrm{pert}} \over 2 \pi} \ln \left|
\frac{32}{\varepsilon \lambda} \sin \left( \frac{\lambda}{2} \ln
\frac{8}{ |\varepsilon| \lambda} \right) \right|
\label{TLsd}
\end{equation}

\noindent \citep{S07}.

The parameter $\lambda = | \Omega | / \omega_0$ measures the
separation of the perturbing and guiding resonances in the units
of one quarter of the width of the guiding resonance, because the
resonances separation in frequency space is equal to $\Omega$,
while the guiding resonance width is equal to $4 \omega_0$
\citep{C79}. Therefore, $\lambda$ can be regarded as a kind of the
resonance overlap parameter. In the asymptotic limit of the
adiabatic (slow) case the resonances in the multiplet strongly
overlap, while in the asymptotic limit of the ``fast'' case the
resonances are segregated. However note that the border $\lambda
\approx 1/2$ \citep{S08a} between slow and fast chaos does not
coincide with the borderline between the cases of overlapping and
non-overlapping of resonances: the latter borderline lies much
higher in $\lambda$. For example, in the phase space of the
standard map the integer resonances start to overlap (on
decreasing $\lambda$) at $\lambda \approx 2 \pi / 0.97 \approx
6.5$ \citep{C79}.

\section{The resonance Hamiltonian}
\label{rh}

\cite{Q03,Q09}, based on the dynamical model of \cite{Co75,Co88},
constructed a Hamiltonian of the motion in the solar neighborhood
by adding the perturbations due to the bar and spiral arms to the
unperturbed Hamiltonian of the motion. Namely, Quillen's model
describes interaction of the bar's 2:1 outer Lindblad resonance
with the spiral's 2:1 or 4:1 inner Lindblad resonance. The
resulting Hamiltonian has the form

\vspace{-3mm}

\begin{equation}
H = A[j^2 + \delta j + \beta j^{1/2} \cos \phi + \epsilon j^{1/2}
\cos(\phi + \nu t - \gamma)]
\label{hg}
\end{equation}

\noindent (\citealt[equation~(23)]{Q03}). Here, $j$ and $\phi$ are
the conjugate momentum and phase variables; $A \approx -5.7$;
$\delta$, $\beta$, $\epsilon$, $\nu$, and $\gamma$ are free
unitless parameters. The ranges for numerical values of the
parameters were estimated in \cite{Q03} from observational
physical and kinematical considerations.

The frequency $\nu$ is counted in units of $\Omega$, and,
accordingly, time is in units of $\Omega^{-1}$; here $\Omega$ is
the rotation rate of the epicyclic center. Therefore, one time
unit is equal to one Galactic year at a given distance from the
center of the Milky Way, divided by $2 \pi$.

The first resonant term (that with the coefficient $\beta$) in
equation~(\ref{hg}) corresponds to the perturbation due to the
bar, while the second one (that with the coefficient $\epsilon$)
to the perturbation due to the spiral arms. The strength of the
second term is much greater than that of the first one
\citep{Q03}. Therefore, it is natural to perform a time-dependent
shift $\phi = \psi - \nu t + \gamma$ of the origin of the
coordinate system. This shift makes the second resonance
explicitly the guiding one. The resulting Hamiltonian is

\vspace{-3mm}

\begin{equation}
H = A j^2 + (A \delta + \nu) j + A \epsilon j^{1/2} \cos \psi + A
\beta j^{1/2} \cos (\psi - \nu t + \gamma) . \label{hg1}
\end{equation}

\noindent Then we introduce the parameter $\Delta = A \delta +
\nu$ and make a constant shift $j = p - \Delta / (2 A)$, $\psi =
\varphi - \gamma$, reducing equation~(\ref{hg1}) to the form

\vspace{-3mm}

\begin{equation}
H = A p^2 + A \epsilon \left(p - \frac{\Delta}{2A}\right)^{1/2}
\cos \varphi + A \beta \left(p - \frac{\Delta}{2A}\right)^{1/2}
\cos (\varphi - \nu t) , \label{hg2}
\end{equation}

\noindent and expand the coefficients in the neighborhood of
$p=0$, leaving only the lowest order (constant) terms. This gives

\vspace{-3mm}

\begin{equation}
H = A p^2 + A \epsilon \left(- \frac{\Delta}{2A}\right)^{1/2} \cos
\varphi + A \beta \left(- \frac{\Delta}{2A}\right)^{1/2} \cos
(\varphi - \nu t) . \label{hg3}
\end{equation}

\begin{figure}
\centering
\includegraphics[scale=1.2, bb = 0 0 300 300]{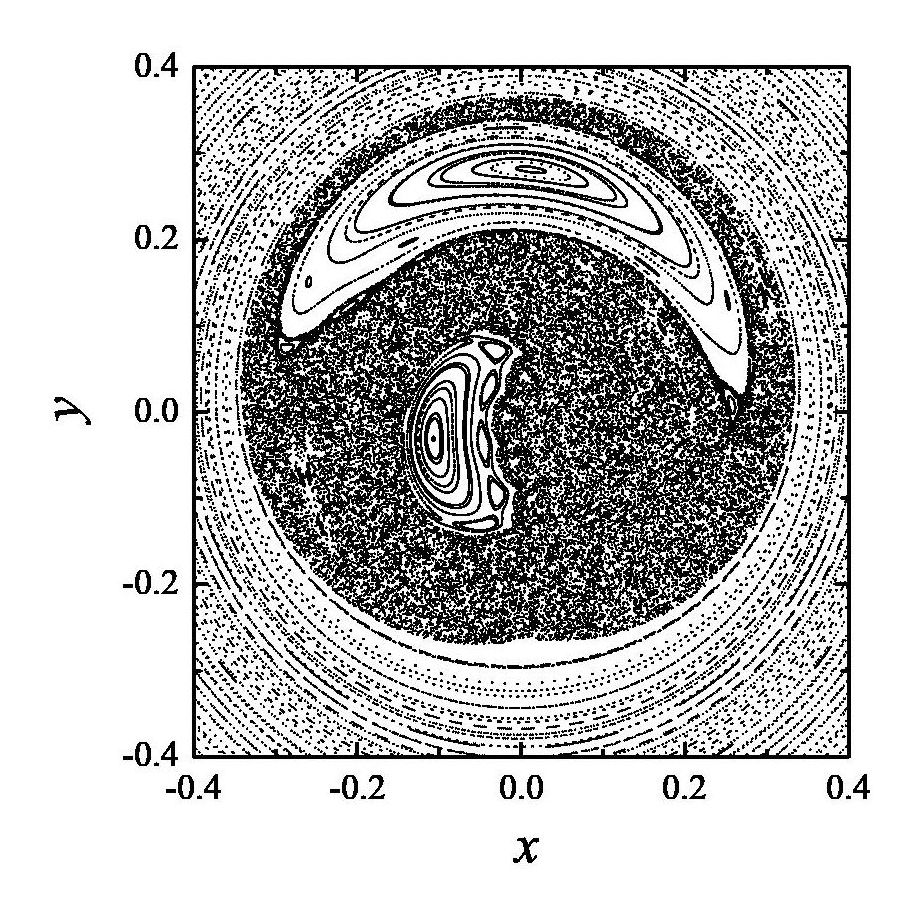}
\caption{Typical example of the phase space section (model~6 of
group~C). The inner regular island (small crescent) corresponds to
the bar's resonance and the outer regular island (big crescent)
corresponds to the spiral's resonance.}
\label{ps_section}
\end{figure}

\begin{figure}
\centering
\includegraphics[scale=1.2, bb = 0 0 300 300]{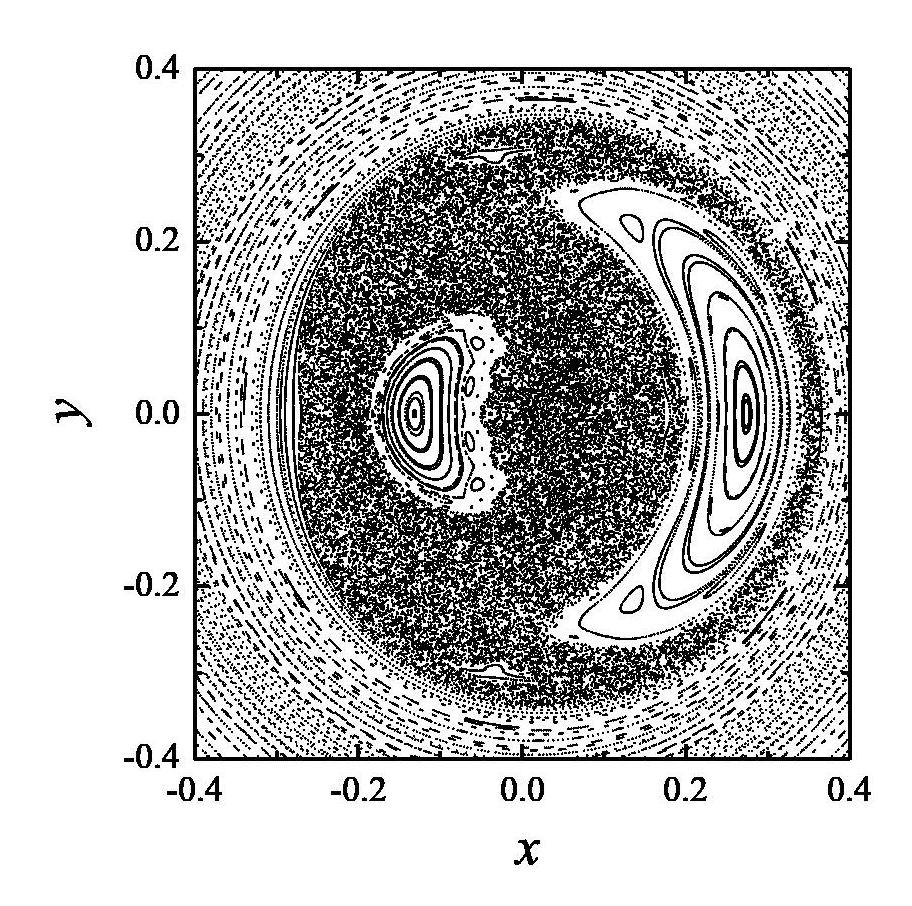}
\caption{Same as in Figure~\protect{\ref{ps_section}}, but
$\gamma = 0$. Apart from the change of the relative angular
orientation of the regular islands, the phase space structure
qualitatively remains the same.}
\label{ps_section_b}
\end{figure}

\noindent Thus we have reduced the perturbed ``second fundamental
model for resonance'' (called so by \citealt{HL83}), given by the
Hamiltonian~(\ref{hg}), to the perturbed ``first fundamental model
for resonance'', given by the Hamiltonian~(\ref{hg3}), i.e., to
the perturbed pendulum model. This has turned out to be possible
because the guiding resonance (that emerging due to the spiral
arms) is bifurcated in all cases, the librational ``crescent'',
born from the bifurcation, being situated always quite far from
the origin of the coordinate system; see the phase space sections
in figures~2--4 in \cite{Q03}, figure~3 in \cite{Q09}, and our
Figures~\ref{ps_section} and \ref{ps_section_b}. The phase space
sections in Figures~\ref{ps_section} and \ref{ps_section_b} have
been constructed by numerical integration of the equations of
motion with the Hamiltonian~(\ref{hg}) in the same way as
described in \cite{Q03}; the coordinates in the sections are $x =
(2 j)^{1/2} \cos \phi$ and $y = (2 j)^{1/2} \sin \phi$.

In Figure~\ref{ps_section}, the phase space section for model~6 of
group~C in \cite{Q03} is shown. It corresponds to figure~4~(f) in
\cite{Q03}, except that we take here a much denser grid of initial
data for the trajectories. In the case of Figure~\ref{ps_section},
the value of $\gamma$ is equal to $\pi/2$, as in \cite{Q03}, while
in the case of Figure~\ref{ps_section_b} the value of $\gamma$ is
chosen to be equal to zero. In both figures, the central regular
island (small crescent) corresponds to the bar's resonance, while
the outer regular island (big crescent) corresponds to the
spiral's resonance. Setting $\gamma = 0$ results mostly in a shift
in the angular location of the spiral's resonance in $\phi$, that
is why the relative angular orientation of the regular islands
changes. The angular orientation of the center of the outer island
(the fixed point of the spiral's resonance) changes by $\pi/2$.
Apart from the change of the relative angular orientation of the
regular islands, the phase space structure is qualitatively one
and the same in Figures. The value of $\gamma$ does not influence
the Lyapunov and diffusion timescales, because this parameter can
be removed by a simple canonical transformation, namely, a
constant shift in time.

The chaotic domain is pronounced in both Figures. It emerges due
to interaction of the bar's and spiral's resonances. The estimates
of the Lyapunov and diffusion times are made in what follows just
for such domains, emerging due to interaction of the resonances,
in various model sets.

Equation~(\ref{hg3}) describes the resonance duad, and, for
estimating the Lyapunov time of the motion in the chaotic layer,
we can apply formulas (\ref{TLfd}) and (\ref{TLsd}) in the cases
of fast and slow chaos, respectively.

\section{Lyapunov time estimates}
\label{lte}

Comparing the Hamiltonians~(\ref{h}) and (\ref{hg3}), one has:
${\cal F} = - A \epsilon \left( - \frac{\delta}{2} - \frac{\nu}{2
A} \right)^{1/2}$, ${\cal G} = 2 A$, $\Omega = \nu$, $a = | A |
\beta \left( - \frac{\delta}{2} - \frac{\nu}{2 A} \right)^{1/2}$,
$b = 0$; $\omega_0 = | A | | \varepsilon |^{1/2} \left( -2 \delta
- \frac{2\nu}{A} \right)^{1/4}$, $\lambda = \frac{| \nu
|}{\omega_0}$, and $\varepsilon = - \frac{\beta}{\epsilon}$. The
results of calculation of the Lyapunov times by formula
(\ref{TLfd}) (when $\lambda > 1/2$, i.e., in the majority of
cases) and by formula (\ref{TLsd}) (when $\lambda < 1/2$, only in
two cases present) are given in Tables~1, 2, and 3 for model
groups~A, B, and C, respectively. The model groups~A, B, and C
correspond to the sets of parameter values considered for the
construction of the phase space sections in figures~2--4 in
\cite{Q03}.

The Hamiltonian~(\ref{hg}) has five parameters: $\delta$, $\beta$,
$\epsilon$, $\nu$, and $\gamma$. All of them are expressed by
means of formulas given in \cite{Q03} through original dynamical
characteristics of the Galaxy and the solar neighborhood, such as
the pattern speeds of the bar and spiral structure, the radius of
the solar circle from the Galactic center, the radius at which the
bar ends, etc. These five parameters elude straightforward
interpretation in terms of the original dynamical characteristics,
but they have straightforward meaning in the framework of the
dynamical model~(\ref{hg}): the parameters $\beta$ and $\epsilon$
characterize the strength of the bar's and spiral's resonances,
respectively; $\delta$ is the ``bifurcation control'' parameter,
as discussed in \cite{Q03}; $\nu$ measures the separation of the
bar's and spiral's resonances in the frequency space; $\gamma$ is
a phase shift, rather unimportant from the dynamical viewpoint, as
discussed at the end of Section~\ref{rh}.

The models 1--10 of each model group correspond to the panel of 10
Poincar\'e sections in a corresponding figure in \cite{Q03}:
group~A corresponds to figure~2, group~B to figure~3, and group~C
to figure~4. The parameters $\beta$ and $\epsilon$ are constants
inside each model group. Their values are given in the notes to
the Tables. The parameters $\delta$ and $\nu$ vary inside each
model group.

The obtained values of the Lyapunov time are all (except in one
case, corresponding to adiabatic chaos) in the range from $\approx
40$ to $\approx 80$ time units. Inspection of the data given in
Tables~1--3 also makes evident that the heuristic estimate $\sim 2
\pi / \nu$ \citep{HM96,Q03} severely underestimates the real
Lyapunov time (by an order of magnitude in many cases). Besides,
it is rather seldom that the considered dynamical systems,
modeling the dynamics in the solar neighborhood, can be called
``fully overlapped'': the values of $\lambda$ (playing the role of
the resonance overlap parameter, see Section~\ref{minr}) are
typically greater than 1/2; so, chaos is non-adiabatic.

Only model~9 (and, marginally, model~8) of model group~B can be
called adiabatic. The values of the problem parameters $\epsilon$,
$\beta$, $\nu$, and $\delta$ in this case are such that the value
of the separatrix map parameter $\lambda$ is less than 1/2.
Generally, the adiabaticity of chaos in any model can be checked
by calculating $\lambda = | \nu | / \omega_0$, given the values of
$\epsilon$, $\beta$, $\nu$, and $\delta$.

As soon as our time unit is equal to one Galactic year divided by
$2 \pi$, we see that the obtained typical Lyapunov times are
approximately equal to 10 Galactic years. How much is it, in
comparison with, e.g., Lyapunov times of the solar system bodies,
in comparable time units? The usual Lyapunov times for asteroids
in the main belt are 3000--10000~years or more \citep{S07}, i.e.,
they are $\sim 1000$ asteroidal orbital periods or more, while the
usual Lyapunov times for highly eccentric comets are of the order
of one cometary orbital period \citep{S07}. So, the Lyapunov times
of the chaotic motion in the considered Galactic model are much
less (by at least two orders of magnitude) than the Lyapunov times
for the main belt asteroids, but an order of magnitude greater
than for the comets, if expressed in adequate time units. In
general terms of the loss of predictability of the motion, the
Galactic dynamical chaos is rather strong.

From inspection of Tables~1--3 one can deduce that the Lyapunov
time depends on the model parameters rather weakly being almost of
the same order in all the models. Thus one can expect that
choosing different values for the model parameters, such as
relative strengths of the bar and spiral structures (e.g.,
choosing the spiral structure to be weaker), would not
qualitatively change the typical Lyapunov time value. However,
note that the change of the model may radically reduce the extent
of the chaotic domain. Then chaos is not ``global'' and so the
probability that the Sun belongs to the chaotic domain of the
phase space might be small.

It is interesting that, as follows from the above estimates of
$T_{\mathrm L}$, the age of the Milky Way measured in its Lyapunov
times is about 5--10. This means that now it is already
practically impossible to restore exact initial conditions for the
stellar dynamics in the solar neighborhood from any observational
data.

\section{Diffusion rates and ballistic flights}
\label{dte}

Let us estimate the characteristic times of chaotic transport
(called the diffusion times, when the diffusive approximation is
used) in the chaotic domain of the phase space in the same model
set of \cite{Q03}. Knowledge of these times will allow one to
judge on the possibility for migration of the Sun from the inner
regions of the Milky Way to its current location. Such a
possibility, arising due to overlapping of resonances in the phase
space, was advocated and studied in detail by \cite{MF10} and
\cite{MFC11}. To estimate the typical diffusion times, we shall
base on the approach developed initially by \cite{CV86,CV89} for
the purposes of studies in cometary dynamics. \cite{C99} employed
a similar approach in a study of the separatrix map dynamics (see
page 11 and, in particular, formulas~(20) and (21) in
\citealt{C99}).

First of all, a reservation should be made that it is only very
approximate that we can consider the chaotic transport in the
problem under study as diffusive. The matter is that the value of
$\lambda$, as follows from Tables~1--3, in the majority of models
is rather low: $\lambda \sim 1$. As explained in
\citep[p.~12--13]{C99}, when $\lambda \sim 1$, ``\dots the layer
width is reduced down to the size of a single kick. \dots Hence,
the diffusion approximation becomes inapplicable. Instead, the
so-called ballistic relaxation comes into play which is much
quicker. In other words, a slow diffusive motion \dots is replaced
now by rapid jumps of a trajectory over the whole layer \dots''.
(A ``single kick'' is the energy increment per iteration of the
separatrix map.) Therefore all the diffusion rate estimates that
we make in this Section should be regarded as extrapolation of
diffusive description. In reality, they represent upper bounds for
the real characteristic times of chaotic transport.

According to \cite{CV86,CV89}, the diffusion rate (in the energy
variable) in the main chaotic layer in the phase space of the
Kepler map\footnote{The Kepler map is a kind of a general
separatrix map, the parabolic motion playing the role of a
separatrix \citep{S10}.} approximately equals the mean (over time)
squared energy increment per iteration of this map. Analogously,
in the case of the ordinary separatrix map~(\ref{sm}), the
diffusion rate (in the energy variable) in the main chaotic layer
in the phase space of the map approximately equals the mean
squared energy increment, i.e., $\langle W^2 \sin^2 \tau_i
\rangle$. Averaging over the interval $0 \leq \tau_i < 2 \pi$, we
find the diffusion rate to be $D_\mathrm{map} \approx W^2/2$.

In the case of the separatrix algorithmic map, the chaotic layer
components corresponding to prograde rotations, retrograde
rotations, and librations of the phase variable should be
considered separately. In the two (prograde and retrograde)
rotation cases the diffusion rate $D_\mathrm{map}$ in the energy
in the map phase space obviously equals $\approx (W^+)^2/2$ and
$\approx (W^-)^2/2$ for the prograde and retrograde rotations,
respectively. Employing the formulas for $\Omega$, $\omega_0$,
$a$, $b$ ($b = 0$), $\lambda$, and $\varepsilon$, given at the
beginning of Section~\ref{lte}, one can calculate the parameters
$W^+$ and $W^-$ of the separatrix algorithmic map~(\ref{sam}). If
$\lambda > 1/2$, the equality $b=0$ implies $|W^-| \ll |W^+|$,
while $a=0$ implies $|W^-| \gg |W^+|$. The component of the
chaotic layer corresponding to reverse (or direct) rotations does
not exist, if $W^-$ (or, respectively, $W^+$) is equal to zero;
its measure is zero. The other circulation component with non-zero
measure is described by the ordinary separatrix map~(\ref{sm})
with the parameters $\lambda$ and $W=W^\pm$ (the non-zero value
among $W^+$ and $W^-$); its extent in $w$ is $\approx \lambda | W
|$ (the half-width of the chaotic layer in the case of fast chaos;
see \citealt{S08a}).

Consider the libration component of the chaotic layer. Then $W^-$
and $W^+$ alternate (replace each other) at each iteration of the
separatrix algorithmic map~(\ref{sam}). It is straightforward to
show \citep{S07} that, if $W^-$ or $W^+$ is equal to zero, the
separatrix algorithmic map~(\ref{sam}) on the doubled iteration
step reduces to the ordinary separatrix map~(\ref{sm}) with the
doubled value of $\lambda$ and the value of $W$ equal to a
non-zero value of $W^\pm$. Taking into account that one iteration
of the new map corresponds to two iterations of the original one,
the diffusion rate referred to the original map time units is

\begin{equation}
D_\mathrm{map} \simeq \frac{1}{4}(W^\pm)^2 ,
\label{Dmapgen}
\end{equation}

\noindent where $W^\pm$ is the non-zero value among $W^+$ and
$W^-$.

For the circulation component of the layer, one has
$D_\mathrm{map} \simeq (W^+)^2/2$ (if $b = 0$) or $D_\mathrm{map}
\simeq (W^-)^2/2$ (if $a = 0$).

As soon as the libration motion is reducible to the ordinary
separatrix map~(\ref{sm}) with the doubled value of $\lambda$, the
layer's extent in $w$ on the side of librations doubles, it
becomes $\approx 2 \lambda | W |$. Note that the parameters
$\lambda$ and $W$ are considered here as independent from each
other. Therefore, the chaotic domain corresponding to libration
dominates in extent, and for a rough estimate of the diffusion
rate over the entire layer it is sufficient to make an estimate
for the libration component alone.

In the case of the Hamiltonian~(\ref{hg3}) $b=0$ and $\eta=0$, so,
$W^\pm = W^+$, where

\begin{equation}
W^+ = \varepsilon \lambda A_{2}(\lambda) = 4 \pi \varepsilon
\lambda^2 {\exp({{\pi \lambda} / 2}) \over \sinh (\pi \lambda)}
\label{Wp}
\end{equation}

\noindent (see equation~(\ref{Wplus})), therefore

\begin{equation}
D_\mathrm{map} \simeq \left( 2 \pi \varepsilon \lambda^2
{\exp({{\pi \lambda} / 2}) \over \sinh (\pi \lambda)} \right)^2
\label{Dmap}
\end{equation}

\noindent in the libration case.

To obtain the diffusion rate referred to real time units, it is
necessary to transform the map time units into the real ones. This
is performed by dividing the diffusion rate referred to map time
units by the mean period of phase rotations (half-periods of
librations) inside the chaotic layer, because this mean period is
nothing but the average time interval corresponding to one map
iteration. Consequently, the diffusion rate referred to real time
units is $D = | \Omega | D_\mathrm{map} / T_{\mathrm{sx}}$, where
$T_{\mathrm{sx}}$ is given by formula~(\ref{Tsx}).

We define the characteristic diffusion time across the chaotic
layer to be equal to the inverse of the diffusion rate. Therefore,
it is just the time needed for the diffusing particle to cover the
relative energy interval equal to one. Note that the maximum
possible deviation in the relative energy $w$ from zero in the
libration case is equal to $-2$ \citep{C79}; therefore, the
defined diffusion time gives an appropriate time estimate for the
global mixing inside the chaotic layer, of course, if the chaotic
layer is broad enough.

In our Hamiltonian~(\ref{hg3}) $b = 0$, so, $W^\pm = W^+$, and one
gets for the diffusion time

\begin{equation}
T_{\mathrm{d}} = \frac{1}{D} =
\frac{T_{\mathrm{sx}}(\lambda, W^+)}{| \Omega | D_\mathrm{map}} \simeq
\frac{4 T_{\mathrm{sx}}(\lambda, W^+)}{| \Omega | (W^+)^2} ,
\label{Td}
\end{equation}

\noindent where

\begin{equation}
T_{\mathrm{sx}}(\lambda, W^+) \approx
\lambda \ln {32 e \over \lambda | W^+ |}
\label{Tsxp}
\end{equation}

\noindent (see equation~(\ref{Tsx})), $e$ is the base of natural
logarithms, and $W^+$ is given by formula~(\ref{Wp}). Finally one
has

\begin{equation}
T_{\mathrm{d}} \simeq
\frac{4 \lambda}{| \Omega | (W^+)^2} \ln {32 e \over \lambda | W^+ |} .
\label{Tdp}
\end{equation}

In the case of the prograde rotation component of the chaotic
layer, the diffusion rate $D_\mathrm{map} \simeq (W^+)^2/2$;
therefore, the diffusion time is two times less than that given by
formula~(\ref{Tdp}).

The results of the calculation of the diffusion times
$T_{\mathrm{d}}$ by formula~(\ref{Tdp}) are given in Tables~4, 5,
and 6 for model groups~A, B, and C, respectively.

Inspection of the data in Tables~4--6 makes it evident that the
diffusion times vary considerably inside the model groups: by as
much as four orders of magnitude in group~C. Taking into account
that our time unit is equal to one Galactic year divided by $2
\pi$, i.e., our time unit $\approx 32$~Myr (1~Myr $= 10^6$~year; 1
Galactic year $\approx 200$~Myr), it is clear that the obtained
diffusion times in most of the models with ordinal numbers up to
three are greater than 10~Gyr (1~gigayear $\equiv$ 1~Gyr $=
10^9$~year), making large-scale radial chaotic migration
improbable. Such ``junior'' models all have positive values of
$\delta$. However, models~7--9 in group B also have large values
of $T_{\mathrm d}$.

\cite{MF10} found in detailed numerical experiments that, due to
the resonance overlap of the bar and spiral structure, the
Galactic disk mixes in about 3 Gyr. From our Tables 4--6 it is
obvious (taking into account that our time unit $\approx 32$~Myr)
that most of the models with negative values of $\delta$ provide
chaotic mixing effective enough to be consistent with the
numerical-experimental results of \cite{MF10}. Almost all models
with large ordinal numbers, except models~7--9 in group B, permit
such migration. Generally, as follows from data in Tables~4--6,
the diffusion rate strongly depends on the radial position in the
Galaxy.

Note that the Hamiltonian model~(\ref{hg}) assumes that the
guiding radius in the unperturbed dynamics is fixed. If migration
is present, the guiding radius varies; so, a more refined model
should be used to account for this variation self-consistently.
One should also mention that chaotic transport can be ubiquitous
in the Galaxy, though its dynamical origin is presumably different
from that considered here. \cite{QDB10} showed that the diffusion
might occur in many regions due to interaction of multiple waves,
and so overlapping of resonances should be common; besides, if the
bar slows down (e.g., \citealt{WK07} and references therein)
resonances can sweep through the Galaxy \citep{QDB10}. Of course,
these processes cannot be described in the specific dynamical
model considered here, but this model already provides an insight
into the possible effectiveness of chaotic transport in the
Galaxy.

The most important conclusion of this Section is that models that
permit large-scale radial chaotic migration of the Sun (from the
inner regions of the Milky Way to its current location) do exist.
This confirms the dynamical possibility of the migration concept
advocated by \cite{MF10} and \cite{MFC11}. What is more, due to
the possibility of ballistic flights mentioned in the beginning of
this Section, the chaotic mixing might be far more effective and
quicker than in the case of normal diffusion. Obviously, the
effect of ballistic relaxation should be explored in detail in the
future.

\section{Discussion and conclusions}

We have considered how the Lyapunov and diffusion timescales of
the stellar dynamics in the solar neighborhood can be estimated.
We have used Quillen's (2003) model to describe interaction of the
``spiral'' and ``bar'' nonlinear resonances in the phase space of
the motion. A method of analytical estimation of the maximum
Lyapunov exponents of the orbital motion has been applied to the
solar neighborhood dynamics. The analytical treatment has been
performed within a framework of the separatrix map theory
\citep{S00a,S02,S07}, describing the motion near the separatrices
of a perturbed nonlinear resonance.

The Lyapunov times turn out to be basically in the range from 6 to
13 Galactic years. In comparison with the Lyapunov times of the
solar system bodies (made in adequate time units), the Galactic
dynamical chaos is rather strong in general terms of the loss of
predictability of the motion. An interesting inference is that, as
soon as the age of the Milky Way measured in its Lyapunov times is
about 5--10, now it is already practically impossible to restore
exact initial conditions for the stellar dynamics in the solar
neighborhood from any observational data.

We have estimated also the diffusion times, based on the approach
developed initially by \cite{CV86,CV89} for the purposes of
studies in cometary dynamics. We have found that, in a number of
models, the diffusion times turn out to be small enough to permit
radial chaotic migration of the Sun from the inner regions of the
Milky Way to its current location. In other words, dynamically
adequate models that permit large-scale radial chaotic migration
do exist. This confirms the dynamical possibility of the migration
concept advocated by \cite{MF10} and \cite{MFC11}. Due to the
possibility of ballistic flights inside the chaotic layer, arising
because $\lambda \sim 1$, the chaotic mixing might be even far
more effective and quicker than in the case of normal diffusion.

We have shown that only in a narrow range of possible values of
the problem parameters $\epsilon$, $\beta$, $\nu$, and $\delta$
the Galactic chaos is adiabatic because the values of the
separatrix map parameter $\lambda$, playing the role of the
resonance overlap parameter, are typically greater than 1/2; in
other words, adiabatic chaos ($\lambda < 1/2$) seems to be not
characteristic for the dynamics in the solar neighborhood.

The author is thankful to Alice Quillen for advice and comments.
It is also a pleasure to thank the referee for helpful remarks.

\newpage

\begin{table}[h!]
\caption{Lyapunov time estimates for model group~A
($\epsilon =-0.004$, $\beta =0.0006$; $\varepsilon =0.15$; this model group
corresponds to the graph panel in figure~2 in \protect\cite{Q03})}
\vspace{-8mm}
\footnotesize
\begin{center}
\begin{tabular}[t]{|c|c|c|c|c|c|c|c|c|c|c|}
\hline
Model & 1 & 2 & 3 & 4 & 5 & 6 & 7 & 8 & 9 & 10 \\
\hline
$\delta$ &    $0.068$ & $0.034$ & $0.018$ & $0.001$ & $-0.006$ & $-0.009$ & $-0.016$ & $-0.019$ & $-0.032$ & $-0.049$ \\
\hline
$\nu$ &       $1.000$ & $0.750$ & $0.625$ & $0.500$ & $0.450$ & $0.425$ & $0.375$ & $0.350$ & $0.250$ & $0.125$ \\
\hline
$\lambda$       & $4.07$ & $3.13$ & $2.65$ & $2.15$ & $1.94$ & $1.84$ & $1.64$ & $1.53$ & $1.11$ & $0.56$ \\
\hline
$T_{\mathrm L}$ & $44.3$ & $39.7$ & $38.5$ & $38.3$ & $38.8$ & $39.3$ & $40.5$ & $41.5$ & $48.2$ & $74.2$ \\
\hline
\end{tabular}
\end{center}
\end{table}

\begin{table}[h!]
\caption{Lyapunov time estimates for model group~B
($\epsilon =-0.004$, $\beta =0.0005$; $\varepsilon =0.125$;
this model group corresponds to the graph panel in figure~3 in \protect\cite{Q03})}
\vspace{-8mm}
\footnotesize
\begin{center}
\begin{tabular}[t]{|c|c|c|c|c|c|c|c|c|c|c|}
\hline
Model & 1 & 2 & 3 & 4 & 5 & 6 & 7 & 8 & 9 & 10 \\
\hline
$\delta$ &    $0.068$ & $0.034$ & $0.018$ & $0.001$ & $-0.006$ & $-0.009$ & $-0.016$ & $-0.019$ & $-0.039$ & $-0.066$ \\
\hline
$\nu$ &       $0.700$ & $0.465$ & $0.347$ & $0.230$ & $0.183$ & $0.159$ & $0.112$ & $0.089$ & $-0.052$ & $-0.240$ \\
\hline
$\lambda$       & $3.37$ & $2.32$ & $1.78$ & $1.20$ & $0.97$ & $0.85$ & $0.60$ & $0.48$ & $0.29$ & $1.42$ \\
\hline
$T_{\mathrm L}$ & $49.4$ & $46.8$ & $49.0$ & $57.1$ & $64.3$ & $69.9$ & $87.5$ & $69.6$ & $123.7$ & $60.4$ \\
\hline
\end{tabular}
\end{center}
\end{table}

\begin{table}[h!]
\caption{Lyapunov time estimates for model group~C
($\epsilon =-0.002$, $\beta =0.0006$; $\varepsilon =0.3$; this model group
corresponds to the graph panel in figure~4 in \protect\cite{Q03})}
\vspace{-8mm}
\footnotesize
\begin{center}
\begin{tabular}[t]{|c|c|c|c|c|c|c|c|c|c|c|}
\hline
Model & 1 & 2 & 3 & 4 & 5 & 6 & 7 & 8 & 9 & 10 \\
\hline
$\delta$ &    $0.068$ & $0.034$ & $0.018$ & $0.001$ & $-0.006$ & $-0.009$ & $-0.016$ & $-0.019$ & $-0.032$ & $-0.066$ \\
\hline
$\nu$ &       $1.200$ & $0.840$ & $0.660$ & $0.480$ & $0.408$ & $0.372$ & $0.300$ & $0.264$ & $0.120$ & $-0.240$ \\
\hline
$\lambda$       & $6.44$ & $4.78$ & $3.89$ & $2.95$ & $2.55$ & $2.35$ & $1.93$ & $1.72$ & $0.82$ & $2.01$ \\
\hline
$T_{\mathrm L}$ & $76.4$ & $60.4$ & $53.2$ & $47.4$ & $46.1$ & $46.0$ & $47.0$ & $48.6$ & $74.1$ & $60.9$ \\
\hline
\end{tabular}
\end{center}
\end{table}

\begin{table}[h!]
\caption{Diffusion time estimates for model group~A}
\vspace{-4mm}
\footnotesize
\begin{center}
\begin{tabular}[t]{|c|c|c|c|c|c|c|c|c|c|c|}
\hline
Model & 1 & 2 & 3 & 4 & 5 & 6 & 7 & 8 & 9 & 10 \\
\hline
$\lambda$ &       $4.07$ &  $3.13$ &  $2.65$ &  $2.15$ &  $1.94$ &  $1.84$ &  $1.64$ &  $1.53$ &  $1.11$ &  $0.56$ \\
\hline
$W^+$ &           $0.104$ & $0.271$ & $0.412$ & $0.595$ & $0.673$ & $0.709$ & $0.771$ & $0.798$ & $0.813$ & $0.506$ \\
\hline
$T_{\mathrm d}$ & $7900$ &  $1100$ &  $440$  &  $210$   & $160$ &   $140$  &  $120$  &  $120$  &  $120$ &  $400$ \\
\hline
\end{tabular}
\end{center}
\end{table}

\begin{table}[h!]
\caption{Diffusion time estimates for model group~B}
\vspace{-4mm}
\footnotesize
\begin{center}
\begin{tabular}[t]{|c|c|c|c|c|c|c|c|c|c|c|}
\hline
Model & 1 & 2 & 3 & 4 & 5 & 6 & 7 & 8 & 9 & 10 \\
\hline
$\lambda$ &       $3.37$ &  $2.32$ &  $1.78$ &  $1.20$ &  $0.97$ &  $0.85$ &  $0.60$ &  $0.48$ &  $0.29$ &  $1.42$ \\
\hline
$W^+$ &           $0.179$ & $0.442$ & $0.608$ & $0.687$ & $0.646$ & $0.600$ & $0.451$ & $0.358$ & $0.200$ & $0.681$ \\
\hline
$T_{\mathrm d}$ & $3000$ &  $450$  &  $240$  &  $210$  &  $250$   & $310$ &   $610$ &   $1000$ &  $4100$ &  $230$ \\
\hline
\end{tabular}
\end{center}
\end{table}

\begin{table}[h!]
\caption{Diffusion time estimates for model group~C}
\vspace{-4mm}
\footnotesize
\begin{center}
\begin{tabular}[t]{|c|c|c|c|c|c|c|c|c|c|c|}
\hline
Model & 1 & 2 & 3 & 4 & 5 & 6 & 7 & 8 & 9 & 10 \\
\hline
$\lambda$ &       $6.44$           & $4.78$ &  $3.89$ &  $2.95$  & $2.55$ &  $2.35$ & $1.93$ & $1.72$ & $0.82$ & $2.01$ \\
\hline
$W^+$ &           $0.013$          & $0.094$ & $0.253$ & $0.638$ & $0.893$ & $1.04$ & $1.35$ & $1.50$ & $1.41$ & $1.30$ \\
\hline
$T_{\mathrm d}$ & $9.4 \times 10^5$ & $13000$ & $1600$ &  $230$   & $110$   & $84$   & $49$   & $41$   & $60$   & $70$ \\
\hline
\end{tabular}
\end{center}
\end{table}

\newpage

\end{document}